\documentclass[preprints,article,accept,pdftex,moreauthors]{Definitions/mdpi} 
\usepackage{physics}
\usepackage{bm}
\usepackage{amsmath}
\newcommand{\ee}{\mathrm{e}}
\newcommand{\ii}{\mathrm{i}}

\newcommand{\NN}{\mathrm{NN}}
\newcommand{\hard}{\mathrm{hard}}
\newcommand{\soft}{\mathrm{soft}}
\newcommand{\coll}{\mathrm{coll}}
\renewcommand{\AA}{\mathrm{AA}}

\newcommand{\T}{\mathrm{T}}
\newcommand{\E}{\mathrm{E}}
\newcommand{\N}{\mathrm{N}}
\newcommand{\bias}{\mathrm{bias}}
\renewcommand{\min}{\mathrm{min}}
\renewcommand{\max}{\mathrm{max}}
\allowdisplaybreaks
\firstpage{1} 
\pubvolume{1}
\issuenum{1}
\articlenumber{0}
\pubyear{2026}
\copyrightyear{2026}
\datereceived{ } 
\daterevised{ } 
\dateaccepted{ } 
\datepublished{ } 

\Title{Geometric bias and centrality dependence of jet quenching in high-energy nuclear collisions}

\Author{Changle Sun $^{1}$\orcidA{}, Yichao Dang $^{1}$\orcidB{} and Shanshan Cao $^{1,}$*\orcidC{}}

\AuthorNames{Firstname Lastname, Firstname Lastname and Firstname Lastname}

\address{%
$^{1}$ \quad Institute of Frontier and Interdisciplinary Science, Shandong University, Qingdao, Shandong 266237, China
}

\corres{Correspondence: shanshan.cao@sdu.edu.cn}

\abstract{
Jet quenching provides a valuable measure of the opacity of the quark-gluon plasma (QGP) produced in high-energy heavy-ion collisions. However, substantial suppression of charged hadron spectra is observed in highly peripheral collisions, despite the expectation of negligible jet-QGP interactions in this regime. To address this, we develop a HIJING-based initial condition model that accounts for the impact parameter dependence of both inelastic nucleon-nucleon (NN) collisions and the number of hard partonic scatterings per inelastic NN collision. This dependence introduces a geometric bias effect on the jet yield within a given centrality class of nucleus-nucleus (AA) collisions, suppressing the high-transverse-momentum hadron spectrum in peripheral collisions due to dilute nucleon overlap at large AA impact parameters. By combining this improved initial condition model with a linear Boltzmann transport model for jet-QGP interactions, we obtain a satisfactory description of the centrality dependence of charged hadron suppression in Pb+Pb collisions at $\sqrt{s_\mathrm{NN}}=5.02$~TeV. 
}


\begin{document}

\section{\label{sec:introduction}Introduction}

High-energy nuclear collisions at the Relativistic Heavy-Ion Collider (RHIC) and the Large Hadron Collider (LHC) enable the exploration of nuclear matter at extremely high density and temperature. It is now well established that a color-deconfined state of matter, the quark-gluon plasma (QGP), is created in relativistic heavy-ion collisions, which exhibits properties of a strongly coupled fluid~\cite{Busza:2018rrf,Elfner:2022iae,Harris:2024aov}. Two main signatures of the QGP are the collective flow of low-energy (soft) particles emitted from the QGP~\cite{Ollitrault:1992bk,Heinz:2013th} and the yield suppression of the high-energy (hard) particles~\cite{Wang:1992qdg,Qin:2015srf,Majumder:2010qh,Cao:2020wlm,Wang:2025lct,Mehtar-Tani:2025rty}. The latter is known as jet quenching, caused by the nuclear modification of hard partons produced from the initial hard scatterings. Over the past two decades, studies on collective flow and jet quenching have transitioned from qualitative discovery into quantitative characterization of the QGP. Key properties of the QGP, such as its shear and bulk viscosities~\cite{Bernhard:2019bmu,JETSCAPE:2020shq,Karmakar:2023ity}, equation of state~\cite{Pratt:2015zsa,Liu:2023rfi}, speed of sound~\cite{CMS:2024sgx}, and jet transport coefficient~\cite{JET:2013cls,JETSCAPE:2021ehl,Xie:2022ght}, have been extracted from soft and hard observables.

Interestingly, collective flow is observed not only in large systems created by Au+Au and Pb+Pb collisions, but also in small systems from proton-proton ($pp$), proton-nucleus ($p$A), and deuteron-nucleus ($d$A) collisions~\cite{CMS:2010ifv,ATLAS:2012cix,ALICE:2012eyl,PHENIX:2013ktj,STAR:2015kak,STAR:2022pfn}. Conversely, no clear signature of jet quenching has been detected at these small scales~\cite{STAR:2014qsy,STAR:2024nwm,PHENIX:2015fgy,CMS:2016xef,CMS:2025jbv,ALICE:2017svf,ATLAS:2022iyq}. Consequently, the formation of QGP in these small systems remains highly debated. To address this, considerable efforts have been dedicated to conducting system-size scans of nucleus-nucleus (AA) collisions, bridging the gap between large and small systems and helping identify the critical threshold for the onset of jet quenching~\cite{Faraday:2025prr,Mazeliauskas:2025clt,Pablos:2025cli,CMS:2018yyx,CMS:2025bta,CMS:2026qef,STAR:2026nfy}. Besides relying on different nuclear species for a system-size scan, one can also analyze different centrality classes within a single collision system. A larger centrality value corresponds to a larger impact parameter, and thus a smaller system size in AA collisions. Nevertheless, a close comparison with experimental data indicates a challenge for theoretical models to accurately describe the centrality dependence of jet quenching~\cite{CMS:2016xef,Cao:2017hhk}---when a model is tuned to describe the nuclear modification factor ($R_\AA$) of high-$p_\T$ hadrons in central collisions, it tends to overestimate the $R_\AA$ in peripheral collisions. This discrepancy needs to be resolved before one can use jets as a reliable tool to probe properties of nuclear matter with small sizes. It has been proposed in Ref.~\cite{Loizides:2017sqq} that even in the absence of jet-medium interactions, $R_\AA$ can be smaller than 1 for high-$p_\T$ hadrons in peripheral AA collisions due to a geometric bias effect. This suppression arises from the larger average nucleon-nucleon (NN) impact parameter ($b_{\mathrm{NN}}$) in peripheral AA collisions compared to unbiased $pp$ collisions, which reduces the probability of hard scatterings per NN inelastic scattering in peripheral AA collisions. This geometric bias effect has been confirmed by measurements of $Z$ bosons~\cite{ATLAS:2019maq,CMS:2021kvd}, where a suppression of their yields is observed in peripheral Pb+Pb collisions at mid-rapidity even though $Z$ bosons do not participate in strong interactions with the QGP. Therefore, alongside improving jet quenching models, it is crucial to incorporate this geometric bias effect into existing models to avoid possible misinterpretation of experimental data in peripheral AA collisions.  

In experiments, the nuclear modification factor of jets (or hadrons) is defined as
\begin{equation}
	R_\mathrm{AA}(p_\T)=\frac{1}{\langle N_\mathrm{coll}\rangle}\frac{\dd N_\AA/\dd p_\T}{\dd N_{pp}/\dd p_\T},
	\label{eq:classicalRAA}
\end{equation}
where $\dd N_\AA/\dd p_\T$ and $\dd N_{pp}/\dd p_\T$ are jet spectra in AA and $pp$ collisions, respectively. The average number of binary (inelastic) NN collisions in one AA collision, $\langle N_\mathrm{coll}\rangle$, is usually estimated using the Glauber model~\cite{Miller:2007ri}. In most calculations, $\dd N_\AA/\dd p_\T$ is obtained with $\langle N_\mathrm{coll}\rangle\, \dd N_{pp}/\dd p_\T \,\otimes\, \text{med}$, where "$\otimes\,\text{med}$" denotes convolution of QGP effects on the jet spectrum. We neglect cold nuclear matter effects on high-energy partons in this work. For convenience, we label this spectrum $\dd N_\AA^\mathrm{med}/\dd p_\T$, and similarly
\begin{equation}
	R_\mathrm{AA}^\mathrm{med}(p_\T)=\frac{1}{\langle N_\mathrm{coll}\rangle}\frac{\dd N_\AA^\mathrm{med}/\dd p_\T}{\dd N_{pp}/\dd p_\T},
\end{equation}
quantifying the QGP effects on jet partons. If one assumes each binary NN collision produces the same number of hard partonic scatterings in AA and $pp$ collisions, $R_\AA^\mathrm{med}$ is the same as $R_\AA$ measured in experiments. This assumption is embedded in the standard Monte-Carlo (MC) Glauber model, where a binary NN collision happens when the transverse distance (or $b_\NN$) between a nucleon pair is smaller than $b^0_\NN=\sqrt{\sigma_\mathrm{in}/\pi}$, with $\sigma_\mathrm{in}$ the inelastic cross section, and each binary NN collision produces one hard partonic scattering in both AA and $pp$ collisions. However, this may not be true in reality. It was suggested in Ref.~\cite{Loizides:2017sqq} that if the number of hard scatterings per binary NN collision depends on $b_\NN$, this number would be different between AA collisions within a given centrality class and unbiased $pp$ collisions, because centrality class biases the average distance between nucleon pairs in AA collisions. This is known as the geometric bias of jet quenching. With this bias taken into account, $\dd N_\AA/\dd p_\T$ should read
\begin{equation}
		\frac{\dd N_\AA}{\dd p_\T}=\frac{\langle N_\AA^\mathrm{hard} \rangle }{\langle N_\NN^\mathrm{hard} \rangle} \frac{\dd N_{pp}}{\dd p_\T} \otimes \text{med} = \frac{\langle N_\AA^\mathrm{hard}\rangle}{\langle N_\coll \rangle \langle N_\NN^\mathrm{hard} \rangle} \frac{\dd N_\AA^\text{med}}{\dd p_\T} \equiv R_\AA^\bias \frac{\dd N_\AA^\text{med}}{\dd p_\T},
		\label{eq:dNAA}
\end{equation}
in which $\langle N_\AA^\mathrm{hard} \rangle$ and $\langle N_\NN^\mathrm{hard} \rangle$ represent the average numbers of hard scatterings in each AA and NN ($pp$) collision, and $R_\AA^\bias \equiv \langle N_\AA^\mathrm{hard} \rangle / ( \langle N_\coll \rangle \langle N_\NN^\mathrm{hard} \rangle )$ is defined as the geometric bias factor. Therefore, the $R_\AA$ measured in experiments and that obtained in most theoretical calculations also differ by this factor:
\begin{equation}
	R_\mathrm{AA}=R_\AA^\bias \times R_\mathrm{AA}^\mathrm{med},
	\label{eq:masterRAA}
\end{equation}
indicating the $R_\AA$ one observes in experiments is not only determined by parton-QGP interactions, but also affected by the initial geometry of AA collisions. 

In this work, we aim to understand the centrality dependence of the charged hadron $R_\AA$ by taking into account effects of both geometric bias and jet-QGP interactions. For geometric bias, we develop an initial condition model based on the Heavy-Ion Jet Interaction Generator (HIJING)~\cite{Wang:1990qp,Wang:1991hta,Gyulassy:1994ew}, in which both the binary NN collision probability and the number of hard partonic scatterings per binary NN collision are dependent on the NN impact parameter. Effects of geometric bias will be illustrated by comparing results from this HIJING-based model to baseline results from the standard MC-Glauber model, where this bias is omitted. Alternative initial condition models exist, such as Angantyr~\cite{Bierlich:2018xfw}, GLISSANDO~\cite{Broniowski:2007nz,Rybczynski:2013yba}, and the one developed in Ref.~\cite{Jia:2009mq}. They implement even more sophisticated NN collision profiles, but involve additional model parameters and are beyond the scope of our current study. Throughout our discussions, some notations may differ from those in Eq.~(\ref{eq:dNAA}), reflecting variations in the detailed calculation of average values. For jet-QGP interactions, we further develop the Linear Boltzmann Transport (LBT) model~\cite{Cao:2016gvr,Luo:2023nsi,Dang:2026ezw} by improving its stability in small QGP systems where parton energy loss is weak. With these developments on both the initial condition model and the jet quenching model, we will show that a satisfactory description of the charged hadron $R_\AA$ is achieved from central to peripheral Pb+Pb collisions at a center-of-mass energy per nucleon pair ($\sqrt{s_\mathrm{NN}}$) of 5.02~TeV.

\section{HIJING-based initial condition \label{sec:hijing}}

HIJING~\cite{Wang:1990qp,Wang:1991hta,Gyulassy:1994ew} is an MC event generator simulating high-energy hadronic and nuclear collisions in which the impact parameter dependence of NN scattering cross section has been taken into account. In this section, we briefly review how this impact parameter dependence is introduced in HIJING and estimate effects of geometric bias in AA collisions. 

\subsection{Nucleon-nucleon collisions}

In the high-energy limit, the normalized thickness function of a nucleon $T_\N(r_\perp)$ can be obtained from the proton electric form factor \(G_\E(\bm{q}_\perp^2)\) via
\begin{equation}
	T_\N(r_\perp) = \frac{1}{(2\pi)^2} \int \dd^2\bm{q}_\perp  G_\E(\bm{q}_\perp^2) \ee^{-\ii\bm{q}_\perp \cdot \bm{r}_\perp},
\end{equation}
where $\bm{r}_\perp$ is the transverse coordinate relative to the center of the nucleon, 
and \(G_\E(\bm{q}_\perp^2)\) can be approximated by an empirical dipole function as~\cite{Albrecht:1965ki,Brodsky:1976rz,Brodsky:1976mn}:
\begin{equation}
	G_\E(\bm{q}_\perp^2) \approx \left[ 1 + \frac{\left|\bm{q}_\perp\right|^2}{\nu^2} \right]^{-2}, \quad \nu^2 = 0.71  (\text{GeV})^2.
\end{equation}
Therefore, one can obtain~\cite{Barger:1987id}
\begin{equation}
	T_\N(r_\perp) = \frac{\nu^2}{4\pi} \nu r_\perp K_1(\nu r_\perp),
\end{equation}
together with the overlap function between two nucleons as
\begin{equation}
\begin{aligned}
	T_{\NN}(b_\NN) &= \int \dd[2]{\bm{r}_\perp}  T_\N(|\bm{b}_\NN - \bm{r}_\perp|)T_\N(\bm{r}_\perp)\\
	&= \frac{\nu^2}{12\pi} \frac{1}{8} (\nu b_\NN)^3 K_3(\nu b_\NN),
	\label{eq:T_{NN}_def}
\end{aligned}
\end{equation}
with $\bm{b}_\NN$ the impact parameter between the two nucleons, and $K_{1/3}$ the Bessel function of the first/third kind. Later, we will only take the form of $T_{\NN}(b_\NN)$ above, but re-adjust the parameter $\nu$ to fit the inelastic cross section of NN collisions at a given beam energy.

For a $\bm{b}_\NN$ between two nucleons, the average number of hard partonic scatterings is given by $\langle N_\NN^\mathrm{hard}({b}_\NN)\rangle=\sigma_{\mathrm{hard}}T_\NN(b_\NN)$, with $\sigma_\mathrm{hard}$ the hard partonic scattering cross section that is responsible for jet production. If hard collision happens for a particular NN event in HIJING, the actual number ($j$) of hard collisions in this event is then sampled from a truncated Poisson distribution as:
\begin{equation}
	g_j(b_\NN)=\frac{[\sigma_{\mathrm{hard}}T_\NN(b_\NN)]^j}{j!}\ee^{-\sigma_{\mathrm{hard}}T_\NN(b_\NN)}, \quad j\geq 1. \label{eq:sjet3}
\end{equation}
On the other hand, the probability of having at least one soft interaction without any hard process is given by
\begin{align}
	\label{eq:sjet4}
	g_0(b_\NN) &= \left[1 - \ee^{-\sigma_{\mathrm{soft}}T_\NN(b_\NN)}\right] \ee^{-\sigma_{\mathrm{hard}}T_\NN(b_\NN)} \\
	&= \ee^{-\sigma_{\mathrm{hard}}T_\NN(b_\NN)} - \ee^{-(\sigma_{\mathrm{soft}} + \sigma_{\mathrm{hard}})T_\NN(b_\NN)}, \nonumber
\end{align}
in which $\sigma_\mathrm{soft}$ denotes the soft NN scattering cross section that is responsible for inelastic scatterings without jet production. The total inelastic NN scattering cross section then reads
\begin{equation}
	\begin{aligned}
		\sigma_{\rm{in}} &= \int \dd[2]{\bm{b}_\NN} \sum_{j=0}^{\infty} g_j(b_\NN) \\
		&= \int \dd[2]{\bm{b}_\NN} \left[ 1 - \ee^{-(\sigma_{\mathrm{soft}} + \sigma_{\mathrm{hard}})T_\NN(b_\NN)} \right]. 
		\label{eq:in_sigma}
	\end{aligned}
\end{equation}
By comparing the equation above with the inelastic cross section obtained from the eikonal formalism~\cite{glauber1959lectures,Proceedings:1974eka,Wang:1990qp}
\begin{equation}
	\sigma_{\mathrm{in}} = \int \dd[2]{\bm{b}_\NN} \left[ 1 - \ee^{-2\chi(b_\NN,s)} \right],
	\label{eq:cin2}
\end{equation}
one can extract the eikonal function as
\begin{equation}
\begin{aligned}
	\chi(b_\NN,s) &= \frac{1}{2} \sigma_{\mathrm{soft}}(s) T_\NN(b_\NN,s) + \frac{1}{2} \sigma_{\mathrm{hard}}(s) T_\NN(b_\NN,s) \\
	&\equiv \chi_{\mathrm{soft}}(b_\NN,s) + \chi_{\rm{hard}}(b_\NN,s),
	\label{eq:eiko}
\end{aligned}
\end{equation}
in which $s$ represents the center-of-mass energy of an NN collision.

To satisfy the geometrical scaling properties~\cite{DiasDeDeus:1973lde,Amaldi:1979kd} observed in experiments, e.g., a near constant elastic-to-total-cross-section-ratio ($\sigma_\mathrm{el}/\sigma_\mathrm{tot}$) in the range of $10<\sqrt{s}<100$~GeV, it is suggested that $\chi_{\mathrm{soft}}(b_\NN,s)$, the dominating term in the corresponding energy range, takes the form of $\chi_{\mathrm{soft}}(\xi)$, where 
\begin{equation}
	\xi = \frac{b_\NN}{b_0(s)}, \quad \text{with} \quad \pi b_0^2(s) = \sigma_0(s).
\end{equation}

Inspired by Eq.~(\ref{eq:T_{NN}_def}), the following parametrization is taken in HIJING~\cite{Wang:1990qp}:
\begin{equation}
	T_\NN(b_\NN, s) = \frac{\chi_0(\xi)}{\sigma_0(s)}, 
	\label{eq:newTNN}
\end{equation}
with
\begin{equation}
	\chi_0(\xi)=\frac{\mu_0^2}{96} (\mu_0 \xi)^3 K_3(\mu_0 \xi),
	\label{eq:chi0}
\end{equation}
satisfying $\int_0^\infty \dd \xi^2 \chi_0(\xi)=1$ and thus the normalization of the overlap function $\int \dd^2 \bm{b}_\NN T_\NN(b_\NN, s) =1$. Here, $\mu_0$ is a model parameter to be determined by the NN inelastic cross section soon.

By setting $\sigma_\mathrm{soft}(s)=2\sigma_0(s)$, the soft part of Eq.~(\ref{eq:eiko}) is further reduced to
\begin{equation}
	\chi_\soft(\xi)=\chi_0(\xi),
\end{equation}
and thus
\begin{equation}
	\chi(b_\NN,s) = \chi(\xi,s) = \left[ 1 + \frac{\sigma_{\mathrm{hard}}(s)}{\sigma_{\mathrm{soft}}(s)} \right] \chi_0(\xi).
\end{equation}

Following Ref.~\cite{Wang:1990qp}, an $s$-independent cross section for the soft part is taken as $\sigma_{\mathrm{soft}}(s) = 57$~mb. For NN collisions at $\sqrt{s} = 5.02$~TeV, the cross section for $2\rightarrow 2$  hard partonic scatterings can be calculated using perturbative QCD as $\sigma_{\mathrm{hard}}=166.2 $~mb~\cite{Loizides:2017sqq}. The parameter $\mu_0$ in Eq.~(\ref{eq:chi0}) is then determined to be $4.25$ by requiring Eq.~(\ref{eq:in_sigma}) or~(\ref{eq:cin2}) to reproduce the inelastic cross section of $p+p$ collisions at $\sqrt{s} = 5.02$~TeV---$\sigma_\mathrm{in}=70$~mb~\cite{CMS:2015nfb}.

\begin{figure}[tbp!]
        \centering
        \begin{minipage}{.47\textwidth}
        \centering
	\includegraphics[width=0.98\linewidth]{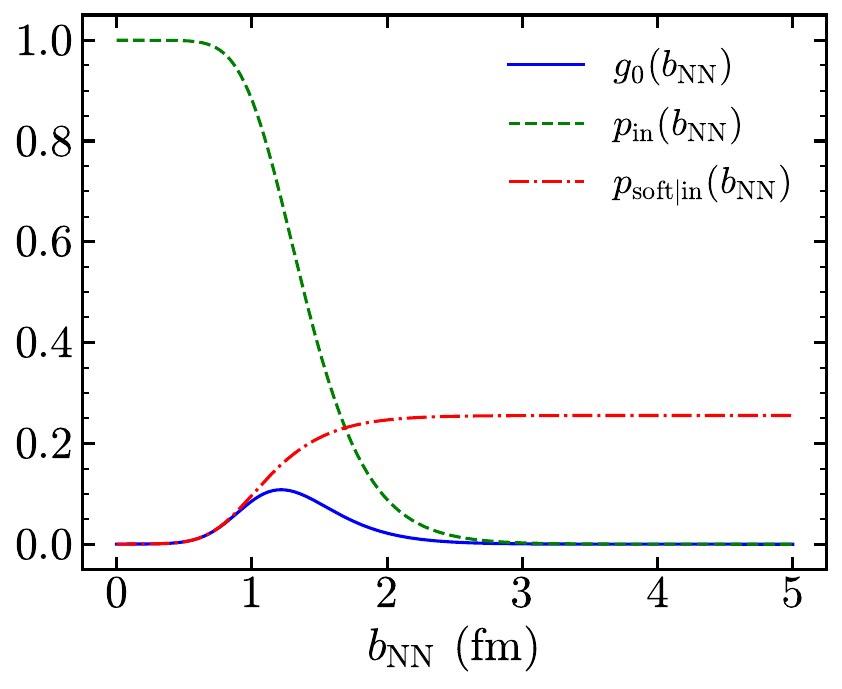}
	\caption{(Color online) Impact parameter dependences of the inelastic NN scattering probability ($p_\mathrm{in}$), the probability of purely soft scattering ($g_0$), and the conditional probability of purely soft scattering given inelastic scattering ($p_\mathrm{soft | in}$) at $\sqrt{s}=5.02$~TeV.}
	\label{fig:probinelastichijing}
	\end{minipage}
	\hspace{0.04\linewidth}
	\begin{minipage}{.47\textwidth}
        \centering
	\includegraphics[width=0.95\linewidth]{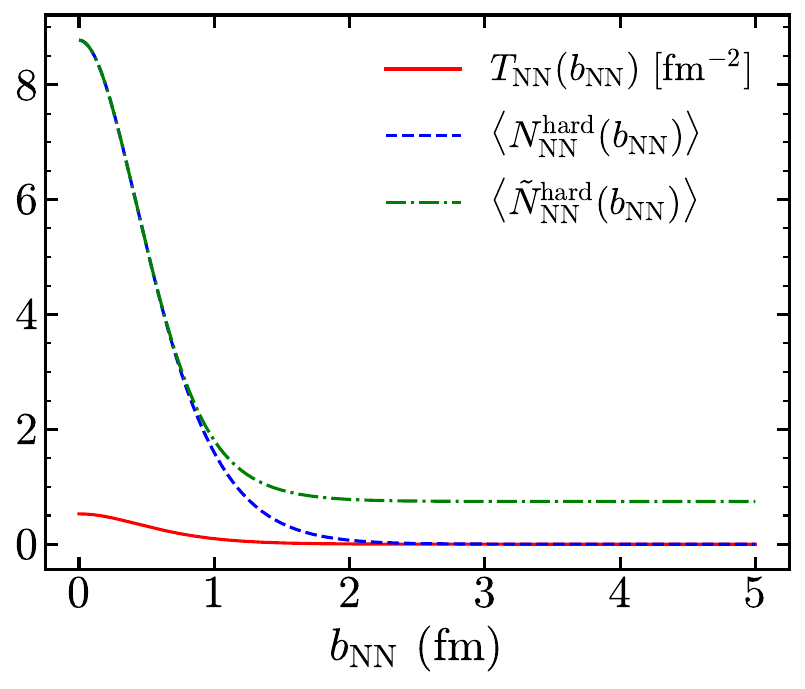}
	\caption{(Color online) Impact parameter dependences of the overlap function ($T_\NN$), the average number of hard scatterings $\langle N_\NN^\hard\rangle$, and the conditional average number of hard scatterings given inelastic scattering $\langle \tilde{N}_\NN^\hard\rangle$ at $\sqrt{s} = 5.02$ TeV.}
	\label{fig:tnnbhijing}
	\end{minipage}
\end{figure}

Using the model parameters above, we first evaluate the probability of an inelastic NN scattering with a given $b_\NN$ as
\begin{equation}
	p_\mathrm{in}(b_\NN)=\frac{\dd{\sigma_{\mathrm{in}}}}{\dd[2]{\bm{b}_\NN}}=1-\ee^{-(\sigma_{\mathrm{soft}}+\sigma_{\mathrm{hard}})T_\NN(b_\NN)}.
\end{equation}
The result is shown in Fig.~\ref{fig:probinelastichijing}, which decreases from 1 to 0 as $b_\NN$ increases. In the figure, we also present the probability of purely soft scattering [$g_0$ in Eq.~(\ref{eq:sjet4})] and the conditional probability of purely soft scattering given that inelastic scattering occurs $p_\mathrm{soft | in} = g_0/p_\mathrm{in}$. At large $b_\NN$, about 25\% of inelastic scatterings are purely soft.

We further present the overlap function $T_\NN$ given by Eq.~(\ref{eq:newTNN}) and the average number of hard partonic scatterings $\langle N_\NN^\mathrm{hard}\rangle=\sigma_{\mathrm{hard}}T_\NN$ as functions of $b_\NN$ in Fig.~\ref{fig:tnnbhijing}. They both decrease toward 0 as $b_\NN$ increases. In addition, we show the conditional average number of hard partonic scatterings given that inelastic scattering occurs---$\langle \tilde{N}_\NN^\hard\rangle=\langle {N}_\NN^\hard\rangle/p_\mathrm{in}$. This number will be used in our MC simulations later. At large $b_\NN$, $\langle \tilde{N}_\NN^\hard\rangle$ approaches 0.75, consistent with the observation in Fig.~\ref{fig:probinelastichijing} that about 25\% of inelastic scatterings are purely soft. 

In contrast to the hard sphere assumption used in the standard MC-Glauber model, where inelastic scatterings only occur within $b^0_\NN=\sqrt{\sigma_\mathrm{in}/\pi}\approx 1.5$~fm, and one hard scattering is always assumed for each inelastic scattering, Figs.~\ref{fig:probinelastichijing} and~\ref{fig:tnnbhijing} show that inelastic scatterings can occur at $b_\NN>b^0_\NN$ and the number of hard scatterings per NN inelastic scattering depends on $b_\NN$. The $b_\NN$-averaged hard scattering number can be evaluated as
\begin{equation}
	\begin{aligned}
		\langle\langle N_\NN^\hard \rangle\rangle &= \dfrac{1}{\sigma_{\mathrm{in}}} \int \dd^2 \bm{b}_\NN \frac{\dd{\sigma_{\mathrm{in}}}}{\dd[2]{\bm{b}_\NN}} \langle \tilde{N}_\NN^\hard (\bm{b}_\NN) \rangle  \\
		&= \dfrac{1}{\sigma_{\mathrm{in}}} \int \dd^2 \bm{b}_\NN \sigma_\hard T_\NN(b_\NN) \\
		&= 2.38. 
		\label{eq:2averageN}
	\end{aligned}
\end{equation}
The $\langle\langle \ldots \rangle\rangle$ notation denotes a double average, performed first over scattering events conditioned on $b_\NN$, and then over $b_\NN$. This can be treated as the average number of hard partonic scatterings per unbiased NN inelastic scattering.

\subsection{Nucleus-nucleus collisions}

In AA collisions, we use the following Woods-Saxon (Fermi) distribution to sample the positions of nucleons inside a nucleus:
\begin{equation}
	\rho(r)=\rho_0\frac{1+w(r/R)^2}{1+\exp\left(\frac{r-R}{a}\right)}.
\end{equation}
For $^{208}$Pb nucleus, we set nucleon density $\rho_0 = 0.17~\mathrm{fm}^{-3}$, radius $R=6.62$ fm and skin thickness $a$ = 0.594 fm. The parameter $w$, which measures the deviation from a spherical shape, is 0. Nucleon-nucleon correlations are approximated by setting the minimum distance between two nucleons as $d = 0.9$~fm, reflecting the repulsive nature of nuclear force at short distance~\cite{Rybczynski:2010ad,Broniowski:2010jd}. With a nucleus-nucleus impact parameter $b$, the centers of the two nuclei (projectile and target) are placed at $(-b/2, 0, 0)$ and $(b/2, 0, 0)$, respectively, in our computational frame. 

In the standard MC-Glauber model, a pair of nucleons, one from projectile and one from target, participate in an inelastic scattering if their transverse distance satisfies $b_\NN < b^0_\NN$, and one hard partonic collision is counted for each inelastic scattering. In contrast, in the HIJING-based model, we use $p_\mathrm{in}(b_\NN)$ presented in Fig.~\ref{fig:probinelastichijing} to determine whether this pair of nucleons scatter inelastically. For a confirmed inelastic scattering, we use $p_\mathrm{soft | in}(b_\NN)=g_0(b_\NN)/p_\mathrm{in}(b_\NN)$ in Fig.~\ref{fig:probinelastichijing} to determine whether this scattering is purely soft or associated with hard collisions. For the latter, a truncated Poisson distribution based on Eq.~(\ref{eq:sjet3}) is used to sample the number of hard partonic collisions---$
\tilde{N}_\NN^\mathrm{hard}(b_\NN)$---for this NN inelastic scattering. In both the standard MC-Glauber model and the HIJING-based model, a pair of nucleons that scatter inelastically contribute to one "binary collision", with the binary collision vertex defined as the mid-point of the pair. Nucleons that undergo inelastic scatterings are termed "participants". For an AA collision, we use $N_\mathrm{coll}$ to denote its number of binary collisions and $N_\mathrm{part}$ for its number of participants. 

\begin{figure}[!tbp]
\centering
\begin{minipage}{.47\textwidth}
\centering
\includegraphics[width=0.98\linewidth]{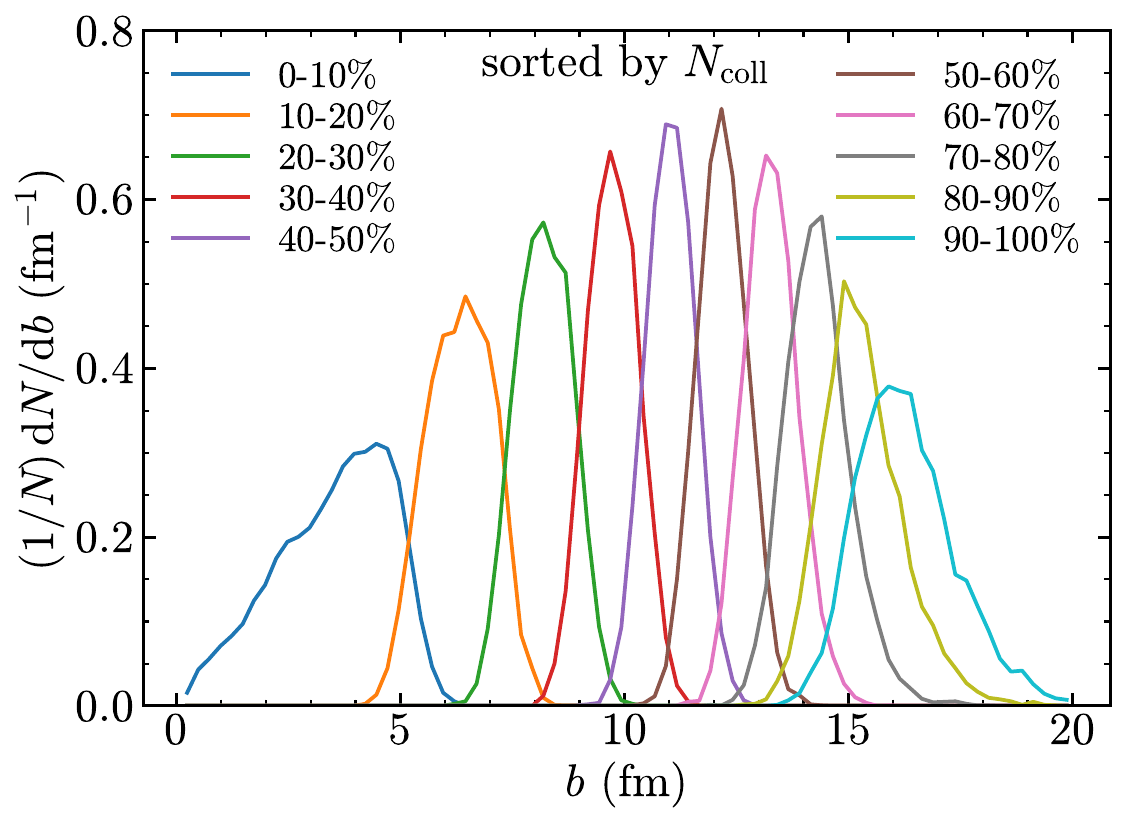}
\end{minipage}
\hspace{0.02\linewidth}
\begin{minipage}{.47\textwidth}
\centering
\includegraphics[width=0.98\linewidth]{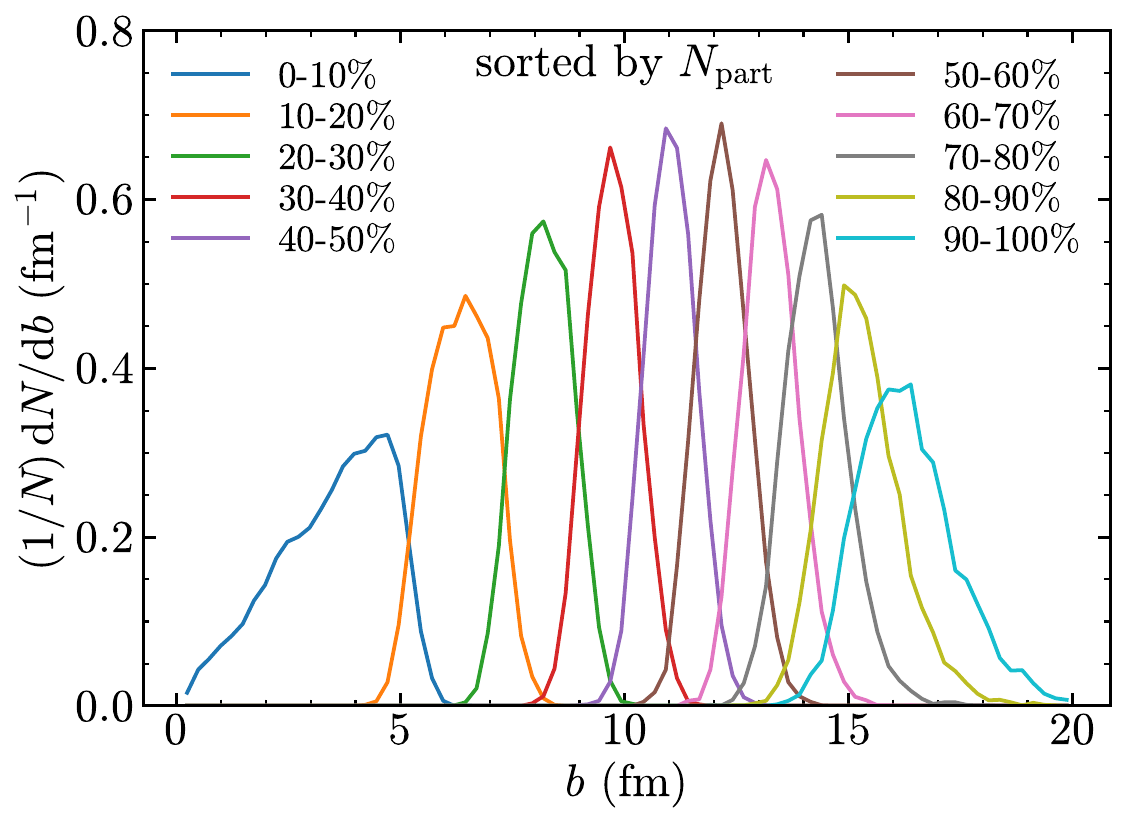}
\end{minipage}
\caption{(Color online) Probability density distributions of the nucleus-nucleus impact parameter $b$ for different centrality classes in Pb+Pb collisions at $\sqrt{s_\mathrm{NN}}=5.02$~TeV, obtained from the HIJING-based initial condition model. Centrality is defined using either $N_\mathrm{coll}$ (left panel) or $N_\mathrm{part}$ (right panel).}
\label{fig:b_pdf_centrality}
\end{figure}

Since the impact parameter ($b$) of AA collisions is not directly measurable, we map it to centrality defined in experiments based on the multiplicities of the final state charged particles. Here, we use the probability distribution of $N_\mathrm{part}$ or $N_\mathrm{coll}$ to approximate that of the charged particle multiplicity. Events are sorted in a descending order of $N_\mathrm{part}$ ($N_\coll$), and the top 10\% events fall into the 0-10\% centrality class, corresponding to central AA collisions with small impact parameters. On the contrary, events at the bottom of this order fall into large centrality classes, corresponding to peripheral collisions with large impact parameters. In Tab.~\ref{tab:centrality_b}, we present the lower and upper boundaries of the impact parameters ($b_\mathrm{min}$ and $b_\mathrm{max}$) for different centrality classes of Pb+Pb collisions at $\sqrt{s_\mathrm{NN}}=5.02$~TeV, obtained using the HIJING-based model. No apparent difference is observed between using $N_\mathrm{part}$ and $N_\coll$ to define centralities. Figure~\ref{fig:b_pdf_centrality} further presents the probability density distributions of the impact parameter $b$ for different centrality classes. A comparison between the left and right panels reveals little difference in these distributions when centrality is defined using $N_\coll$ and $N_\mathrm{part}$.

\begin{table}[htbp]
	\caption{Lower and upper boundaries of impact parameters for different centralities of Pb+Pb collisions at $\sqrt{s_\mathrm{NN}}=5.02$~TeV, determined by the probability distributions of $N_\mathrm{coll}$ and $N_\mathrm{part}$ from the HIJING-based initial condition model.}
	\label{tab:centrality_b}
	\begin{tabularx}{\textwidth}{CCCCC}
		\toprule
		\multirow{2}{*}{\text{Centrality}} & \multicolumn{2}{c}{\text{Sorted by $N_\mathrm{coll}$}} & \multicolumn{2}{c}{\text{Sorted by $N_{\mathrm{part}}$}} \\
		\cmidrule(lr){2-3} \cmidrule(lr){4-5}
		& \text{$b_\min$ (fm)} & \text{$b_\max$ (fm)} & \text{$b_\min$ (fm)} & \text{$b_\max$ (fm)} \\ 
		\midrule
		0-10\%   & 0    & 6.36 & 0    & 6.32 \\
		10-20\%  & 4.10 & 8.42 & 4.13 & 8.49 \\
		20-30\%  & 5.84 & 10.1 & 6.11 & 10.3 \\
		30-40\%  & 7.83 & 11.5 & 7.90 & 11.4 \\
		40-50\%  & 9.22 & 12.8 & 9.28 & 13.1 \\
		50-60\%  & 10.1 & 14.3 & 10.1 & 14.3 \\
		60-70\%  & 11.2 & 16.0 & 11.3 & 16.0 \\
		70-80\%  & 12.2 & 18.3 & 12.1 & 17.9 \\
		80-90\%  & 13.0 & 19.8 & 13.0 & 19.6 \\
		90-100\% & 13.4 & 20.0 & 13.5 & 20.0 \\
		\bottomrule
	\end{tabularx}
\end{table}

\begin{figure}[!tbp]
			\centering
			\includegraphics[width=0.95\linewidth]{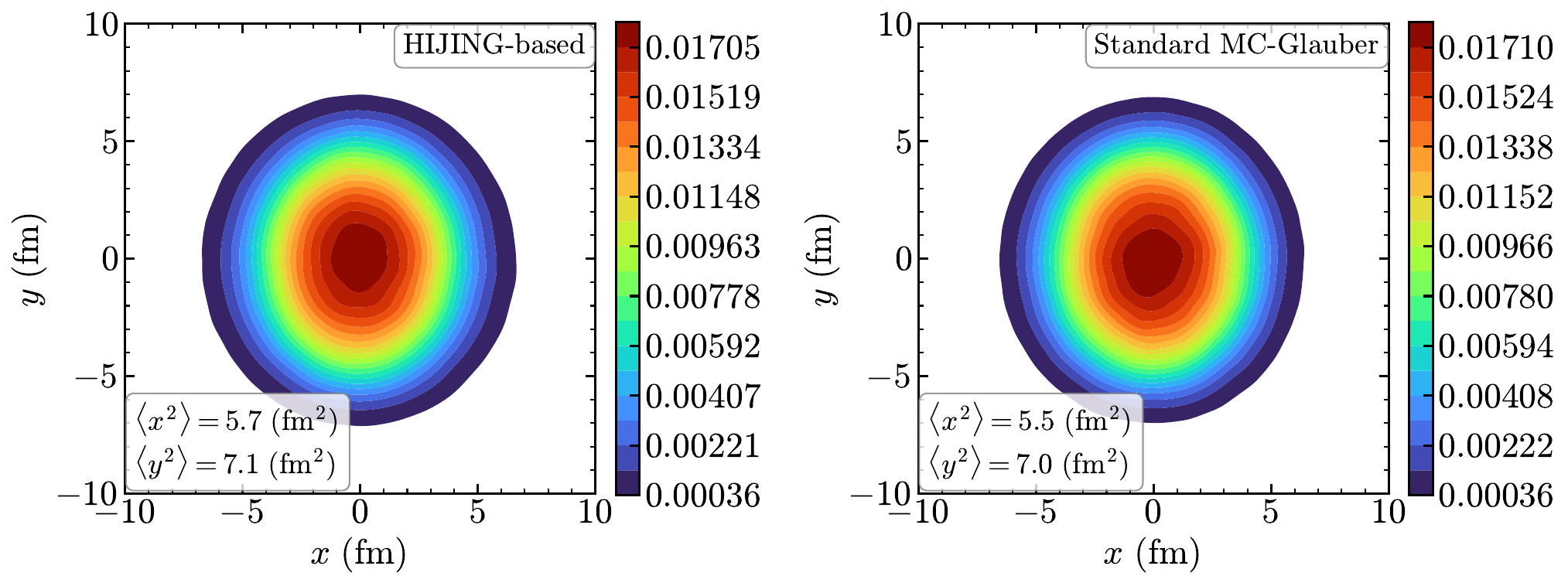}
			\caption{(Color online) Normalized density distributions of hard collision vertices in 0-10\% Pb+Pb collisions at $\sqrt{s_\mathrm{NN}}=5.02$~TeV, compared between the HIJING-based initial condition and the standard MC-Glauber initial condition. }
			\label{fig:kde0010}
		\end{figure}

\begin{figure}[!tbp]
			\centering
			\includegraphics[width=0.95\linewidth]{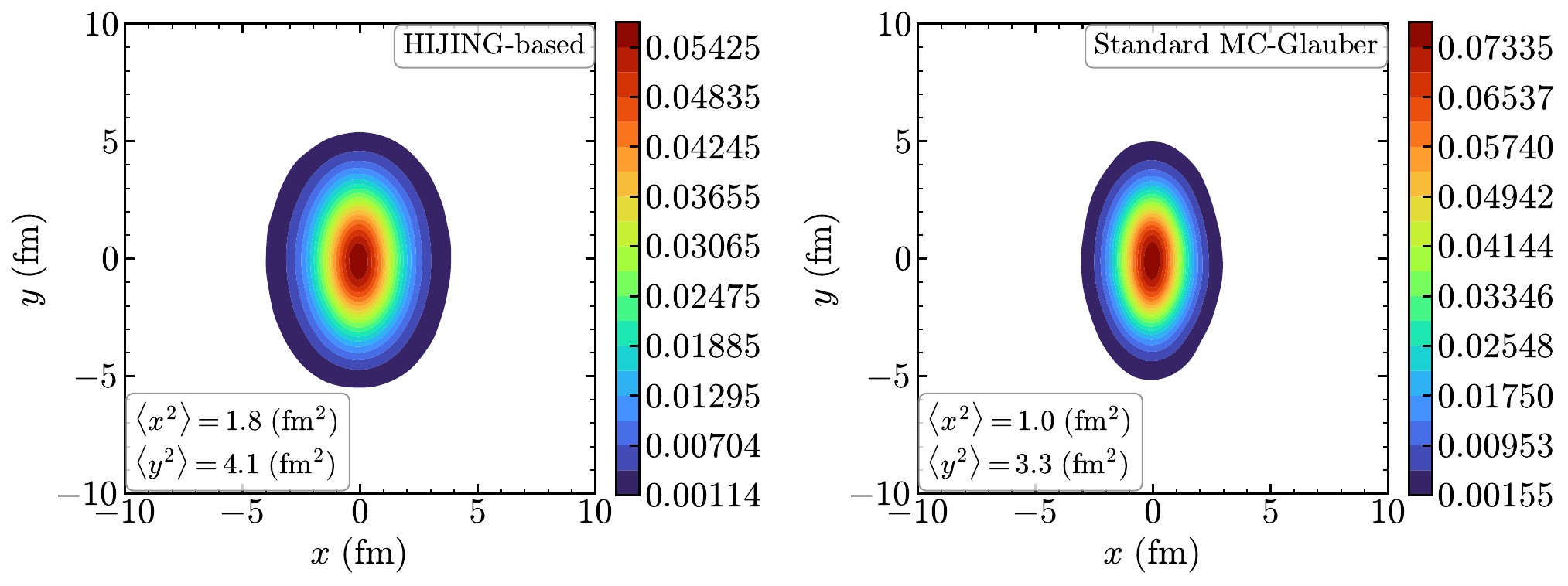}
			\caption{(Color online) Normalized density distributions of hard collision vertices in 50-70\% Pb+Pb collisions at $\sqrt{s_\mathrm{NN}}=5.02$~TeV, compared between the HIJING-based initial condition and the standard MC-Glauber initial condition. }
			\label{fig:kde5070}
		\end{figure}

		In Figs.~\ref{fig:kde0010} and~\ref{fig:kde5070} we present the normalized density distributions of hard collision vertices of Pb+Pb collisions at $\sqrt{s_\mathrm{NN}}=5.02$~TeV for two different centrality bins, 0-10\% and 50-70\%, respectively. For both the standard MC-Glauber model and the HIJING-based model, a hard collision vertex is placed on its corresponding binary collision vertex defined earlier, except that in the former model, one binary collision always generates one hard scattering, while in the latter, one binary collision may generate zero, one, or multiple hard scatterings. No distinct difference between the two models is observed for central collisions in Fig.~\ref{fig:kde0010}. On the other hand, as seen in Fig.~\ref{fig:kde5070}, the hard collision vertices from the HIJING-based model tend to be more sparsely distributed than those from the standard MC-Glauber model in peripheral collisions. This is because the HIJING-based model allows a pair of nucleons far away from each other (a typical configuration in peripheral collisions) to scatter inelastically, while the standard MC-Glauber model only allows them to scatter within $b^0_\NN$. Such different distributions of hard collision vertices can cause less average energy loss of jet partons initialized by the HIJING-based model than that by the standard MC-Glauber model. However, as we will show later, no obvious difference can be seen in the charged hadron $R_\AA$ due to the weak parton energy loss in peripheral collisions.
		
		In our HIJING-based model simulation, for an AA collision with a given $b$ (or within a given centrality bin), its hard collision number is calculated by:
		\begin{equation}
			N^\hard_\AA= \sum_{i=1}^{N_\coll} \tilde{N}_\NN^\hard(b^{\,i}_\NN),
			\label{eq:NAAHard}
		\end{equation}
		where $\tilde{N}_\NN^\hard(b^{\,i}_\NN)$ denotes the number of hard scatterings given that an NN inelastic scattering (indexed by $i$) occurs. The geometric bias factor of jet quenching is then defined as~\cite{Loizides:2017sqq}:
		\begin{equation}
			R^{\bias}_\AA =\dfrac{N^\hard_\AA}{N_\coll \, \langle\langle{N_\NN^\hard}\rangle\rangle},
		\end{equation}
		with $\langle\langle{N_\NN^\hard}\rangle\rangle$ the average number of hard scatterings per unbiased inelastic NN collision, as evaluated earlier in Eq.~(\ref{eq:2averageN}). 
		
		For the standard MC-Glauber model, each binary collision contributes to exactly one hard scattering, leading to $\langle\langle{N_\NN^\hard}\rangle\rangle=1$ and $N^\hard_\AA=N_\coll$, and therefore $R^{\bias}_\AA =1$ for each collision event. In contrast, due to the $b_\NN$-dependence of $N_\NN^\hard$, $R^{\bias}_\AA$ deviates from 1 for the HIJING-based model. This causes a bias in $R_\AA$ that is not related to the nuclear modification of jet partons inside the QGP.
		
\begin{figure}[tbp!]
        \centering
        \begin{minipage}{.47\textwidth}
        \centering
	\includegraphics[width=0.96\linewidth]{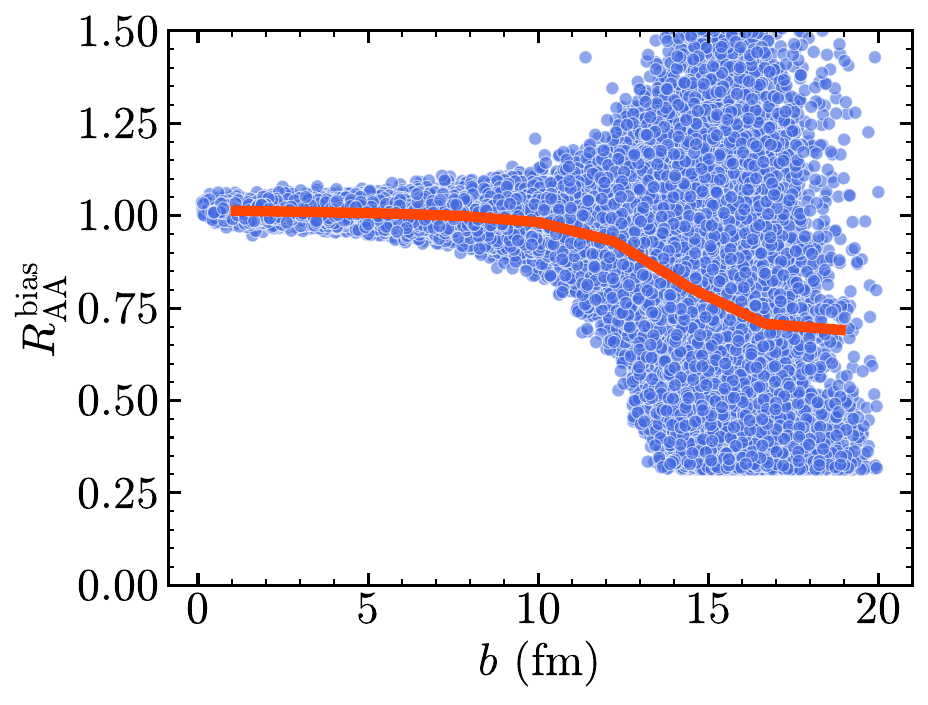}
	\caption{(Color online) The geometric bias factor of jet quenching as a function of nucleus-nucleus impact parameter in Pb+Pb collisions at $\sqrt{s_\mathrm{NN}}=5.02$~TeV, blue dots for individual events and red line for event-averaged value at given impact parameter.}
	\label{fig:raavsb}
	\end{minipage}
	\hspace{0.04\linewidth}
	\begin{minipage}{.47\textwidth}
        \centering
        \vspace{-10pt}
	\includegraphics[width=0.98\linewidth]{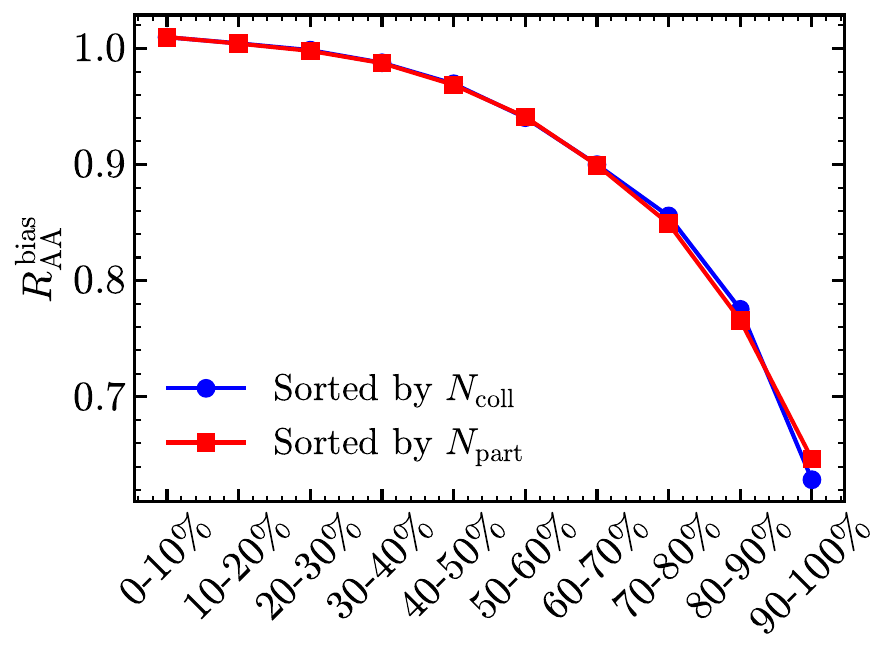}
	\caption{(Color online) The average geometric bias factor of jet quenching in different centrality classes of Pb+Pb collisions at $\sqrt{s_\mathrm{NN}}=5.02$~TeV, compared between $N_\coll$ and $N_\mathrm{part}$ sortings for centrality division.}
	\label{fig:raacentrality}
	\end{minipage}
\end{figure}
		
		In Fig.~\ref{fig:raavsb}, we sample $10^5$ Pb+Pb collision events uniformly distributed in the impact parameter ($\bm{b}$) space with $0<b<20$~fm, and calculate the $R^{\bias}_\AA$ factor for each event, shown as a blue dot in the figure. To suppress the fluctuations of $\tilde{N}^\hard_\NN$ in each binary collision, we use its average value $\langle \tilde{N}^\hard_\NN(b_\NN) \rangle$ from Fig.~\ref{fig:tnnbhijing} in Eq.~(\ref{eq:NAAHard}). Values of blue dots at a given $b$ are further averaged, leading to the average geometric bias factor as a function of $b$, as shown by the red curve. 
		In Fig.~\ref{fig:raacentrality}, we further present the centrality dependence of this $R^\mathrm{bias}_\mathrm{AA}$ factor. The $10^5$ events sampled above are divided into different centrality bins based on either $N_\mathrm{coll}$ or $N_\mathrm{part}$. The average value of $R^\mathrm{bias}_\mathrm{AA}$ is then calculated within each centrality bin using the same method as that used to obtain its $b$ dependence in Fig.~\ref{fig:raavsb}.
		One can clearly observe a "suppression" of jets at large centrality. This "suppression" is not attributed to parton energy loss. Instead, it arises from a smaller hard scattering number per binary collision in peripheral AA collisions compared to unbiased NN collisions, due to the biased (larger) average nucleon-nucleon distance ($b_\NN$) in peripheral AA collisions. As expected, this average $R^{\bias}_\AA$ shows little dependence on whether the centrality bins are defined using $N_\mathrm{coll}$ or $N_\mathrm{part}$ in our initial condition model. Subsequently, we will combine this $R^{\bias}_\AA$ factor with effects of parton energy loss to provide a comprehensive picture of the centrality dependence of the charged hadron $R_\mathrm{AA}$ observed in experiments.

\section{Jet parton production, evolution, and hadronization \label{sec:LBT}}

\subsection{Scale-segmented evolution of jet partons \label{sec:reviewLBT}}

In this work, we employ the Pythia~8~\cite{Sjostrand:2007gs,Bierlich:2022pfr,Sjostrand:2014zea} event generator to simulate the high-$p_\mathrm{T}$ hadron production in $pp$ collisions. This involves processes of the initial hard partonic scatterings that generate high-$p_\mathrm{T}$ partons, their splittings (showers) in vacuum from their initial high virtualities down to the hadronization scale ($Q_\mathrm{h}=0.5$~GeV), and their subsequent string fragmentation into hadrons. In AA collisions, we use Pythia to generate the initial high-$p_\mathrm{T}$ partons, with their spatial distributions determined by either the standard MC-Glauber model or the HIJING-based model described in the previous section. These partons are first evolved down to the medium scale of the QGP ($Q_\mathrm{M}$) via the Pythia vacuum showers. This takes a finite formation time for each parton, evaluated as the sum of its previous splitting times above $Q_\mathrm{M}$~\cite{Zhang:2022ctd}: $\tau_\mathrm{f} = \sum_i 2E_i/Q_i^2$. After that, jet partons interact with the QGP within the linear Boltzmann transport (LBT) model~\cite{Cao:2016gvr,Luo:2023nsi,Dang:2026ezw}. Inside the QGP, their virtualities are held fixed at $Q_\mathrm{M}$. Once the partons exit the QGP, they are returned to Pythia for further vacuum showers down to $Q_\mathrm{h}$, followed by hadronization. The QGP medium is generated by the (3+1)-dimensional CLVisc hydrodynamic model~\cite{Pang:2018zzo,Wu:2021fjf}, with its initial time set as $\tau_0=0.6$~fm and hadronization temperature set as $T_\mathrm{pc}=165$~MeV. Before interacting with the medium, $t< \max (\tau_0, \tau_\mathrm{f})$, jet partons are assumed to stream freely from their production vertices. Below $T_\mathrm{pc}$, jet interactions with the hadron gas are neglected, considering the much more dilute density of the hadron gas compared to the QGP medium.

Here, we briefly summarize the key components of the LBT model and discuss our further developments compared to Ref.~\cite{Dang:2026ezw}. At $Q_\mathrm{M}$ and inside the QGP, the phase space distribution of jet partons ($f_a$) evolves according to the Boltzmann equation as:
\begin{equation}
	\label{eq:boltzmann1}
	p_a \cdot \partial f_a(\bm{x}_a, \bm{p}_a, t) = E_a \left( \mathcal{C}_{\mathrm{el}} + \mathcal{C}_{\mathrm{inel}} \right),
\end{equation}
where $(t,\bm{x}_a)$ is the spacetime coordinate of parton $a$, $p_a=(E_a,\bm{p}_a)$ is its four-momentum, and $\mathcal{C}_{\mathrm{el}}$ and $\mathcal{C}_{\mathrm{inel}}$ are the collision integrals for elastic and inelastic processes, respectively. A collision integral typically consists of a gain term and a loss term, with the latter being related to the scattering rate of a jet parton from a given state into other states. The elastic scattering rate is calculated based on the $2\rightarrow 2$ processes between a jet parton and a medium parton perturbatively at the leading order~\cite{Auvinen:2009qm}. The medium parton is sampled using thermal distributions based on the local temperature of the hydrodynamic medium background. The inelastic scattering rate is related to the number of medium-induced gluon emissions per unit time, with the gluon spectrum taken from the higher-twist energy loss calculation~\cite{Wang:2001ifa,Zhang:2003wk,Majumder:2009ge}. A key parameter in the LBT model is the strong coupling constant $\alpha_\mathrm{s}$, which directly controls the elastic scattering rate via the $2\rightarrow 2$ scattering matrices, and affects the medium-induced gluon spectrum through the jet transport coefficient $\hat{q}$, which characterizes the transverse momentum broadening square per unit time of a jet parton due to its elastic scatterings with the medium. We use a running coupling $\alpha_\mathrm{s}^\mathrm{run}$ for interaction vertices connected to jet partons, and a fixed coupling $\alpha_\mathrm{s}^\mathrm{fix}$ for vertices inside the medium. We set $Q_\mathrm{M}=2$~GeV and $\alpha_{\mathrm{s}}^\mathrm{fix} = 0.34$ in the present study, which was shown to provide reasonable descriptions of various jet quenching data in central (0-10\%) Pb+Pb collisions at $\sqrt{s_\mathrm{NN}}=5.02$~TeV in Refs.~\cite{Dang:2026ezw,Jing:2025bwi}.

In the LBT model, we track and evolve not only jet partons and their emitted gluons, but also the thermal partons that are scattered out of the QGP background. These scattered thermal partons are called "recoil" partons. The production of recoil partons creates particle holes within the QGP, known as back-reaction or "negative" partons, which are also tracked in LBT to ensure energy-momentum conservation of the parton system. Recoil and negative partons constitute "jet-induced medium excitation", which has been shown essential for understanding observables related to fully reconstructed jets~\cite{He:2018xjv,He:2022evt}. For convenience, we collectively refer to jet partons, emitted gluons, and recoil partons as "positive" partons, to distinguish them from negative partons.

\begin{figure}[tbp!]
	\centering
	\includegraphics[width=0.45\linewidth]{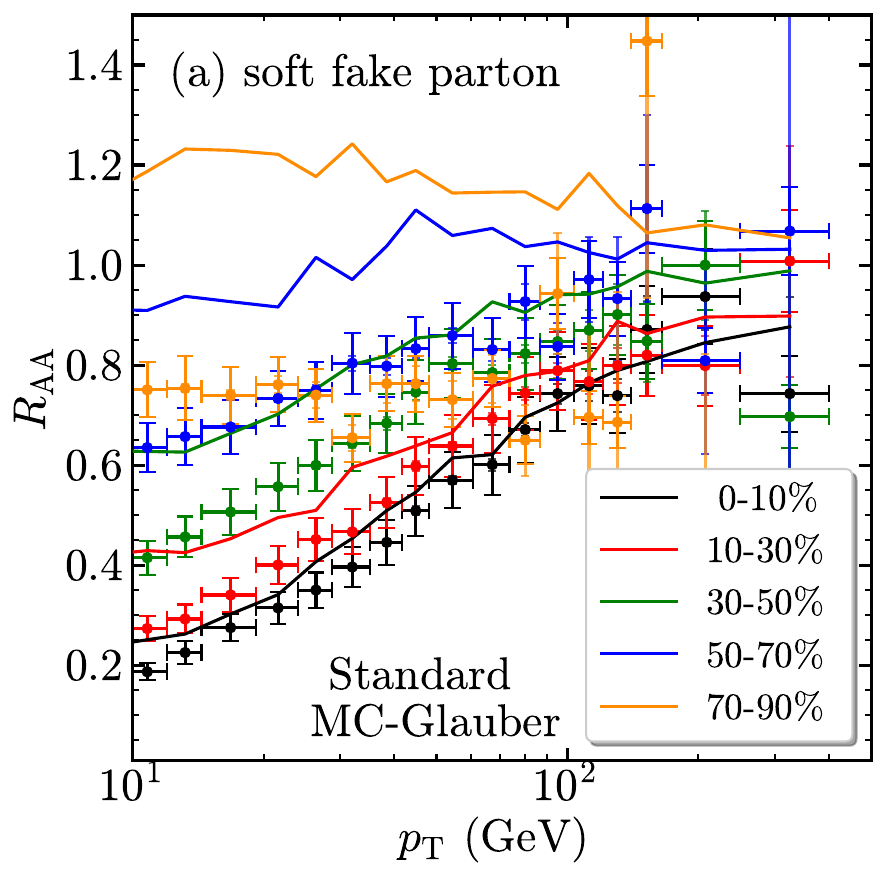}
	\hspace{0.05\linewidth}
	\includegraphics[width=0.45\linewidth]{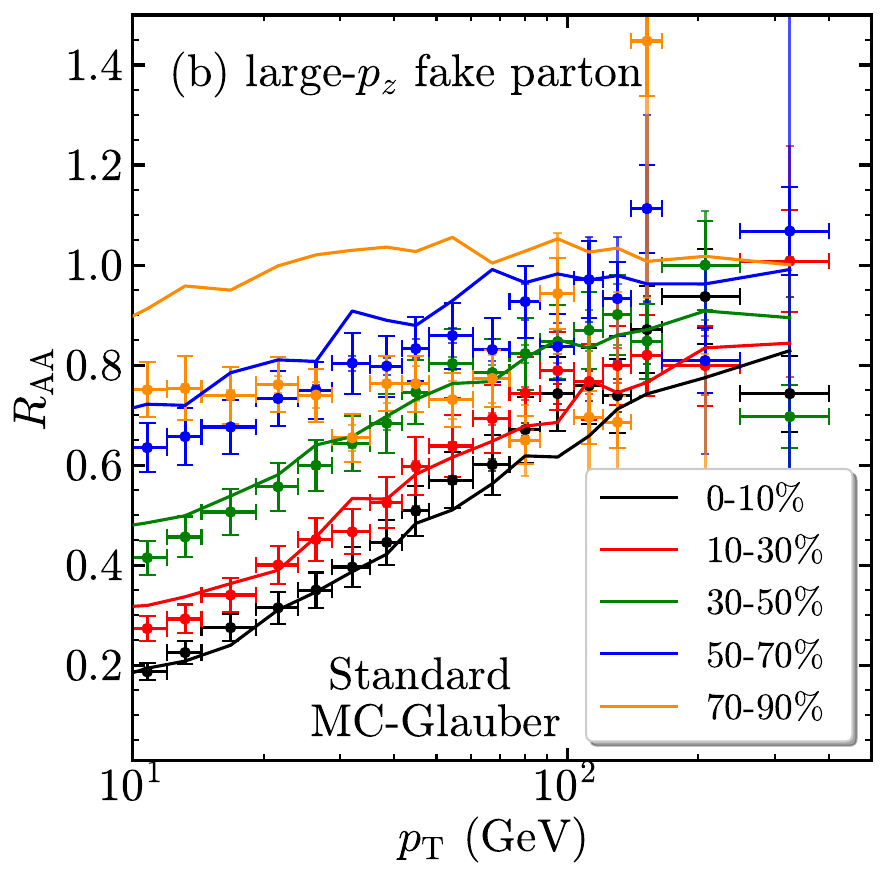}
	\caption{(Color online) The $R_{\AA}$ of charged hadrons in different centrality bins of Pb+Pb collisions at $\sqrt{s_\mathrm{NN}}=5.02$~TeV, (a) from the "soft fake parton" scheme, and (b) from the "large-$p_z$ fake parton" scheme. The standard MC-Glauber model is used for the initial condition of jet production vertices. The CMS data~\cite{CMS:2016xef} are shown for comparison.}
	\label{fig:raa_comparison}
\end{figure}

In order to transfer the final state LBT partons back to Pythia for their further vacuum parton showers and hadronization, the latest LBT model tracks the color information of jet-correlated partons (positive and negative partons) in both elastic and inelastic scatterings inside the QGP. The elastic scattering process can exchange color between a jet parton and a medium parton, and therefore connect the jet parton with a recoil or negative parton into a string. Since negative partons are hypothetical particles that represent energy depletion of the QGP, they cannot be transitioned to Pythia together with positive partons. In Ref.~\cite{Dang:2026ezw}, we replace each negative parton by a "fake" parton, carrying the same color information but a small momentum ($p_x=p_y=p_z=0.1$~GeV). Positive and fake partons are placed together in Pythia for forming positive hadrons through vacuum parton showers and string fragmentation. Meanwhile, negative partons are hadronized separately into negative hadrons using a colorless hadronization model~\cite{JETSCAPE:2019udz,Zhao:2020wcd,Zhao:2021vmu}. This model connects partons into strings based on their momentum-space ordering, without requiring color information. Finally, the negative-hadron contribution to jet observables is subtracted from the positive-hadron contribution.

\subsection{Improvement on the "fake" parton scheme \label{sec:improveFake}}

While the "fake" parton scheme above works well for jet and hadron suppression in central Pb+Pb collisions at $\sqrt{s_\mathrm{NN}}=5.02$~TeV in our earlier work~\cite{Dang:2026ezw}, unphysical results may emerge when we move to smaller QGP systems produced by peripheral Pb+Pb collisions. As shown in Fig.~\ref{fig:raa_comparison}(a), for the 70-90\% centrality bin, the charged hadron $R_\AA$ unexpectedly exceeds 1. Here, $R_\AA$ only includes the medium effect [$R_\AA^\text{med}$ in Eq.~(\ref{eq:masterRAA})] without taking into account the geometric bias factor. We have verified that this $R_\mathrm{AA}$ is exactly 1 when jet-medium interactions are completely switched off in the LBT model. Therefore, this unexpected enhancement likely results from rare scatterings between jet partons and a small QGP system, which cause negligible parton energy loss but alter the color flows of jet partons. Due to these changes in color flow, a high-energy jet parton originally connected to the beam remnants at large rapidity via a string can instead become connected to a low-energy negative (or fake) parton. This affects both the subsequent vacuum parton showers and the hadronization of the high-energy jet parton after it exits the QGP, since these processes rely on the string configuration in Pythia~\cite{Bierlich:2022pfr}.

\begin{figure}[tbp!]
	\centering
	\includegraphics[width=0.5\linewidth]{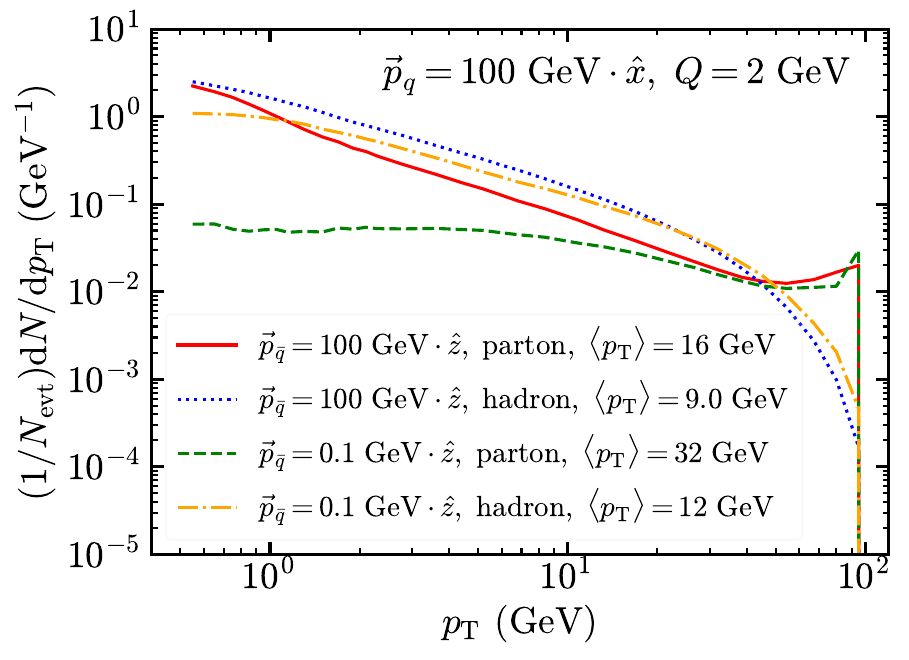}
	\caption{(Color online) The $p_\T$ distributions of partons and hadrons originating from a high-$p_\T$ ($\vec{p}=100$~GeV~$\hat{x}$, $Q=2$~GeV) quark, compared between two string configurations: (1) being connected to a high-$p_z$ (100 GeV) recoiler and (2) being connected to a low-$p_z$ (0.1 GeV) recoiler in the longitudinal direction.}
	\label{fig:p_perp_oredered_shower} 
\end{figure}

In Fig.~\ref{fig:p_perp_oredered_shower}, we investigate how the string configuration affects the final state parton and hadron spectra in Pythia. We start with a high-$p_\T$ quark (100~GeV momentum in the $\hat{x}$ direction) and set its virtuality scale as $Q=2$~GeV. This high-$p_\T$ quark is connected to an antiquark along the longitudinal ($\hat{z}$) direction with either a high (100~GeV) or low (0.1~GeV) momentum. This quark-antiquark pair forms a "dipole" in Pythia, in which the jet parton (quark) is called a "radiator" and its partner (antiquark) is called a "recoiler". In Pythia, the splitting of the radiator is implemented in the rest frame of the dipole system, in which both the radiator and the recoiler are on-shell at first. A portion of the recoiler's energy is shifted to the radiator for raising the radiator's virtuality. With this setup, the splitting of the radiator is simulated based on the parton splitting function $P(z)$ in vacuum, with $z$ the fractional energy taken by a daughter parton from the radiator. Therefore, the momentum of the recoiler in the global frame not only limits the phase space of parton splittings---the maximum energy that can be shifted from the recoiler to the radiator in the dipole rest frame, but also affects the momentum distribution of the final state partons from these splittings in the global frame. As shown in Fig.~\ref{fig:p_perp_oredered_shower}, after we execute the vacuum parton shower routine "forceTimeShower" in Pythia, the final state jet partons tend to maintain higher momenta if a smaller $p_z$ is assigned to the recoiler. This is reflected by both the $p_\T$ distributions and the mean $p_\T$ of partons presented in the figure. Hadronization shifts the particle spectra toward lower $p_\T$, but does not change their relative orderings at the partonic level with respect to the recoiler's $p_z$. As a result, connecting jet partons to low-energy fake partons in jet-medium scatterings enhances the charged hadron $R_\mathrm{AA}$. In central collisions, this enhancement is offset by parton energy loss inside the QGP; while in peripheral collisions, it leads to $R_\mathrm{AA}>1$, as seen in Fig.~\ref{fig:raa_comparison}(a). 

The sensitivity of the final particle spectrum to the recoiler's momentum should not be physical. One may follow the Pythia manual and verify that in a single splitting, if $z$ in $P(z)$ is treated as the light-cone momentum fraction, instead of the energy fraction, taken by a daughter parton from its parent radiator, the final state momentum of the jet parton becomes much less sensitive to the recoiler's momentum. Improving the Pythia kernel for sequential parton splittings is beyond the scope of this study. For the present work, we assign a large longitudinal momentum ($p_z=10^4$~GeV) to each fake parton. This follows the concept implemented in the colorless hadronization model~\cite{JETSCAPE:2019udz,Zhao:2020wcd,Zhao:2021vmu} where the end of a color string is attached to the beam remnants at large rapidity. In the absence of sizable parton energy loss in peripheral collisions, this manipulation reproduces the string configuration of jet partons in $p+p$ collisions and therefore recovers the $R_\AA \sim 1$ baseline. As shown in Fig.~\ref{fig:raa_comparison}(b), by assigning a large $p_z$ to the fake partons, the charged hadron $R_\AA$ becomes consistent with 1 in the 70-90\% centrality bin. Mitigating the unphysical enhancement of the parton spectrum caused by the small recoiler's momentum also improves the centrality dependence of the charged hadron $R_\AA$ relative to that shown in Fig.~\ref{fig:raa_comparison}(a). A simultaneous description of the charged hadron $R_\AA$ is achieved from 0-10\% to 30-50\% centrality bins. Overestimation of $R_\AA$ still exists for centrality regions beyond 50\%, which is caused by the geometric bias effect and will be discussed in the next section. We note that while assigning a large $p_z$ to fake partons improves our model description of the leading hadrons at mid-rapidity, it will introduce artificial enhancement of particle productions at large rapidities. Therefore, it is still necessary to modify the Pythia routine of parton splittings, reducing their sensitivity to the recoiler's momentum, in our future efforts.

\subsection{Hadronization of negative partons\label{subsec:hadronization}}

Another improvement we make relative to the previous work~\cite{Dang:2026ezw} is in the hadronization scheme for negative partons. In Ref.~\cite{Dang:2026ezw}, uncorrelated negative partons produced at different locations inside the QGP are connected into strings that further break into hadrons via the colorless hadronization model~\cite{JETSCAPE:2019udz,Zhao:2020wcd,Zhao:2021vmu}. This implementation can combine low-energy (thermal scale) negative partons into high-energy hadrons and overestimate the negative hadron spectrum around $p_\mathrm{T}=10$~GeV. After subtracting the negative hadron spectrum from the positive one, an over-suppression (even net negative value) of the charged hadron $R_\mathrm{AA}$ is seen around $p_\mathrm{T}=10$~GeV in Ref.~\cite{Dang:2026ezw}, which is unphysical. 

Negative partons should represent the depletion of thermal energy inside the QGP caused by jets. In this work, we use a thermal distribution, $f(\bm{p}) = [\ \text{exp} (\sqrt{\bm{p}^2+m^2}/T_\mathrm{pc})-1\ ]^{-1}$, to sample pions in the local rest frame of the QGP medium, with $m$ set as 0.14~GeV. The energy and momentum of each pion is then boosted to the global frame using the average velocity of negative partons in each event---$\langle\,\vec{v}\,\rangle_\mathrm{neg}=\sum \vec{p}_\mathrm{neg}/ \sum {E}_\mathrm{neg}$. Energy conservation between negative partons and hadrons is enforced by adjusting the energy of the last hadron sampled in each event. Using this hadronization model, we no longer observe the over-suppression of the hadron $R_\AA$ around $p_\mathrm{T}=10$~GeV in Fig.~\ref{fig:raa_comparison}. A more rigorous treatment of the negative partons should be converting them into negative current sources of the hydrodynamic evolution of the QGP, as implemented in the CoLBT-hydro model~\cite{Chen:2017zte}. The particlization of the QGP based on the thermal model will then naturally incorporate the hadron suppression caused by these energy depletion. Nevertheless, the greater computational cost of the CoLBT-hydro model is unnecessary for studying hadron suppression at high-$p_\mathrm{T}$, where the impact of negative partons is negligible.

\section{Effects of geometric bias on hadron suppression \label{sec:results}}

In this section, we investigate the centrality dependence of the charged hadron $R_\AA$ by combining effects of the impact parameter dependence of the hard partonic scattering number in NN collisions discussed in Sec.~\ref{sec:hijing} and parton interactions with the QGP discussed in Sec.~\ref{sec:LBT}. Since a reasonable description of the charged hadron $R_\AA$ from central to semi-peripheral (up to 30-50\% centrality) in Pb+Pb collisions is already achieved in Fig.~\ref{fig:raa_comparison}(b) using the standard MC-Glauber model, and the geometric bias factor is only prominent (over 5\% deviation from 1) beyond 50\% centrality as shown in Fig.~\ref{fig:raacentrality}, we start with investigating geometric effects on the charged hadron $R_\AA$ in peripheral Pb+Pb collisions. 

\begin{figure}[tbp!]
	\centering
	\includegraphics[width=0.48\linewidth]{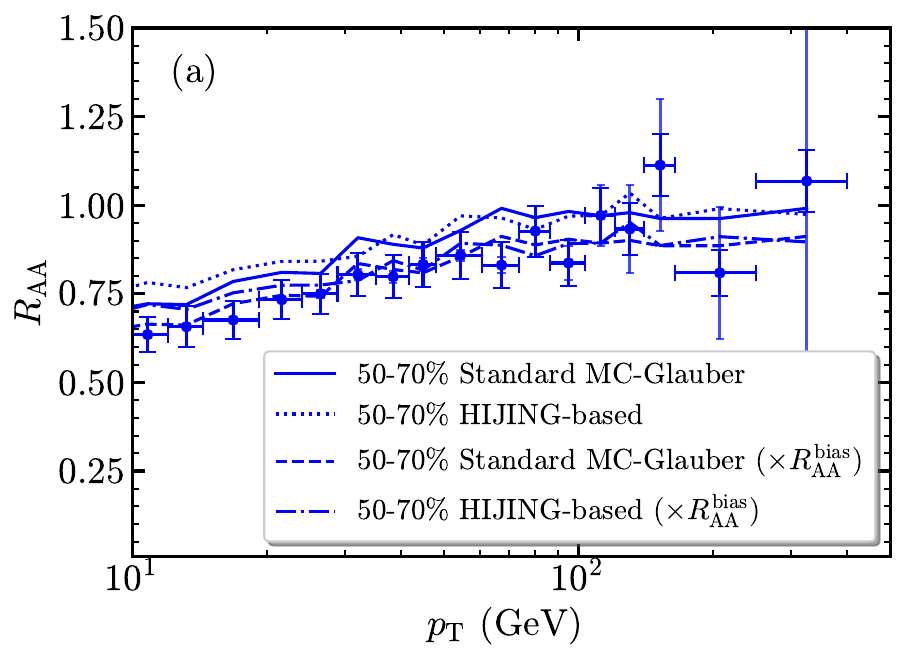}
	\hspace{0.02\linewidth}
	\includegraphics[width=0.48\linewidth]{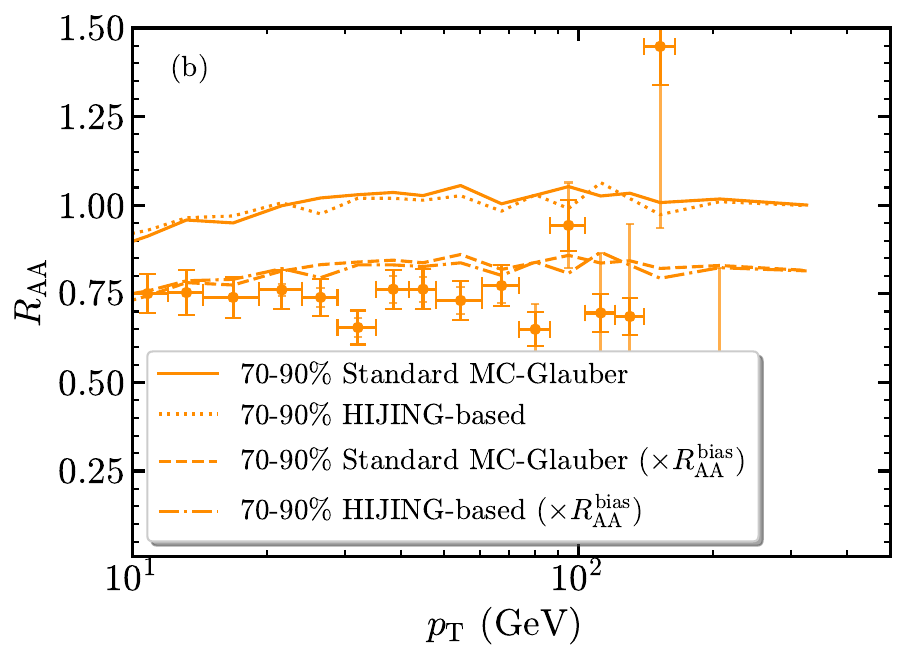}
	\caption{(Color online) The nuclear modification factors of charged hadrons in (a) 50-70\% and (b) 70-90\% Pb+Pb collisions at $\sqrt{s_\mathrm{NN}}=5.02$~TeV, compared between using hard partonic scattering vertices sampled from the standard MC-Glauber model and the HIJING-based model, and between with and without including the geometric bias factor of jet quenching. The CMS data~\cite{CMS:2016xef} are shown for comparison.}
	\label{fig:raa_peripheral_bias}
\end{figure}

Shown in Fig.~\ref{fig:raa_peripheral_bias} are the nuclear modification factors of charged hadrons in (a) 50-70\% and (b) 70-90\% Pb+Pb collisions at $\sqrt{s_\mathrm{NN}}=5.02$~TeV. In each panel, we compare between sampling initial hard scattering vertices using the standard MC-Glauber model and the HIJING-based model, as well as between with and without including the geometric bias factor of jet quenching. In the 50-70\% centrality class, there might be a hint of less suppression (larger $R_\AA$) of charged hadrons initiated from hard scattering vertices drawn from the HIJING-based model than that from the standard Glauber model below $p_\T \sim 30$~GeV. This could result from the smaller parton energy loss caused by the more sparsely distributed initial hard partons from the HIJING-based model in peripheral collisions, as discussed in Fig.~\ref{fig:kde5070}. However, since parton energy loss is weak in peripheral collisions, the difference between the two models is tiny in the 50-70\% centrality class and becomes negligible in the 70-90\% centrality class. Effect of the geometric bias becomes more prominent in more peripheral collisions. Following Eq.~(\ref{eq:masterRAA}), multiplying $R_\AA$ ($R_\AA^\text{med}$) by the bias factor $R_\AA^\mathrm{bias}$ results in a slight decrease of $R_\AA$ for the 50-70\% centrality class, and a significant decrease for the 70-90\% centrality class, agreeing better with the experimental data from the CMS Collaboration~\cite{CMS:2016xef}. This suggests that the experimentally observed suppression of high-$p_\mathrm{T}$ hadrons in highly peripheral heavy-ion collisions is predominantly driven by the initial-state collision geometry rather than the final-state parton energy loss. Without correction for this geometric bias, one would erroneously attribute the observed suppression in peripheral collisions to jet quenching, overestimating the QGP medium effects on jets in small collision systems. We note that the factorization of $R_\AA$ into the geometric bias factor and the nuclear modification effect in Eq.~(\ref{eq:masterRAA}) is based on a separation of time scales. The initial hard collisions happen within the time scale of $1/Q$, with $Q$ the initial virtuality of energetic partons, while parton-QGP interactions occur after $\max(\tau_0, \tau_\mathrm{f})$ as discussed in Sec.~\ref{sec:reviewLBT}, which is later than $1/Q$. This factorization simplifies our calculation of $R_\AA$ with separate evaluations on the geometric bias factor and the medium modification of jet partons.

\begin{figure}[tbp!]
	\centering
	\includegraphics[width=0.45\linewidth]{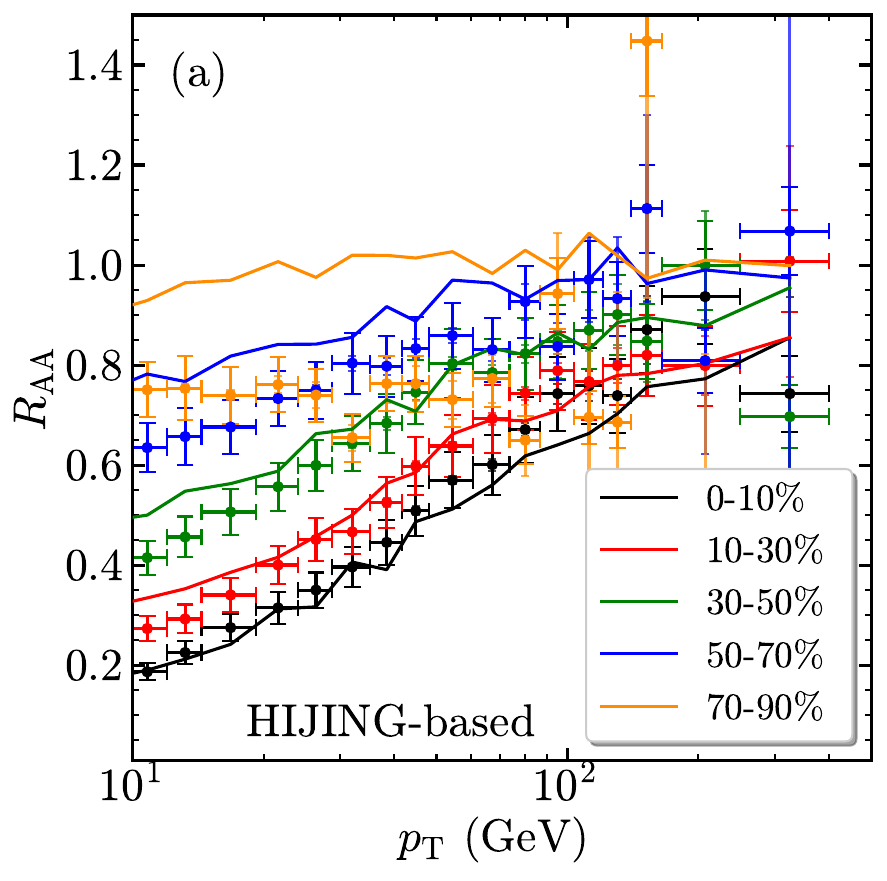}
	\hspace{0.05\linewidth}
	\includegraphics[width=0.45\linewidth]{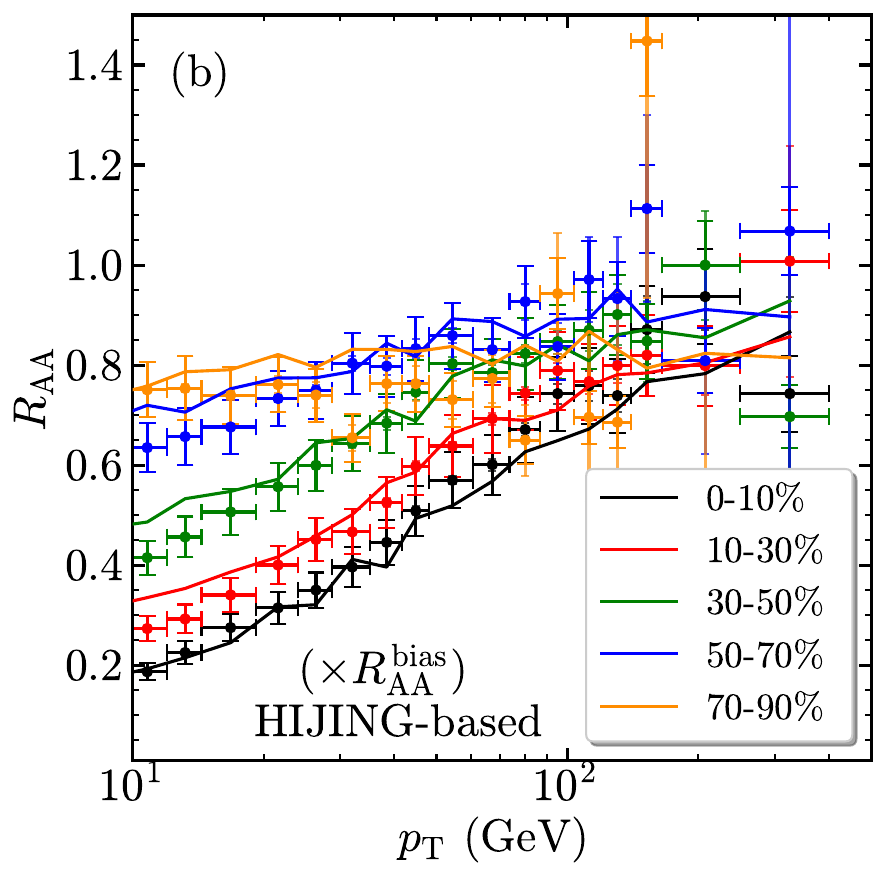}
	\caption{(Color online) The nuclear modification factors of charged hadrons in various centrality classes of Pb+Pb collisions at $\sqrt{s_\mathrm{NN}}=5.02$~TeV, obtained using hard partonic scattering vertices sampled from the HIJING-based model and compared between (a) without and (b) with including the geometric bias factor of jet quenching. The CMS data~\cite{CMS:2016xef} are shown for comparison.}
	\label{fig:raa_hijing_all}
\end{figure}

In the end, we present the $R_\AA$ of charged hadrons based on the HIJING-based initial condition across various centrality classes in Fig.~\ref{fig:raa_hijing_all}, compared between (a) without and (b) with including the geometric bias factor. Comparing between Fig.~\ref{fig:raa_hijing_all}(a) and Fig.~\ref{fig:raa_comparison}(b), we confirm again that the initial vertices of jet partons sampled by the HIJING-based model and the standard Glauber model have little impact on the parton energy loss inside the QGP. Comparing between Fig.~\ref{fig:raa_hijing_all}(a) and Fig.~\ref{fig:raa_hijing_all}(b), one clearly observes the increasing impact of $R_\AA^\text{bias}$ on the charged hadron $R_\AA$ as centrality increases. Overall, by combining the geometric bias effect obtained using the HIJING-based model and the parton-QGP interactions simulated using the LBT model, we achieve a satisfactory description of the centrality dependence of the charged hadron $R_\AA$.

\section{Conclusions\label{sec:conclusion}}

We investigated the centrality dependence of the high-$p_\T$ charged hadron suppression in Pb+Pb collisions at $\sqrt{s_\NN}=5.02$~TeV by disentangling the suppression factor into a geometric bias effect and interactions between jet partons and the QGP. A HIJING-based initial condition was developed to capture the impact parameter dependence of the inelastic NN scattering cross section and the number of hard partonic scatterings per NN inelastic scattering. The LBT model was applied and improved to simulate jet-QGP interactions. In particular, a large $p_z$-scheme was introduced for fake partons (which represent jet-induced energy depletion inside the QGP) in their string connections to jet partons. This eliminates the unphysical enhancement of hadron spectrum in peripheral AA collisions relative to its $pp$ baseline, improving the model's performance for parton interactions with small QGP systems at mid-rapidity.

Compared to the standard MC-Glauber model in which inelastic scattering occurs within a hard cut-off at $b_\NN^0$ and one hard partonic scattering is assigned to each inelastic NN scattering, the HIJING-based model leads to a broader distribution of jet production vertices as well as a smaller $N_\mathrm{coll}$-rescaled hard scattering number ratio between AA and NN collisions in peripheral AA collisions. This ratio is defined as the geometric bias factor, $R_\AA^\text{bias}$, quantifying the difference in the per-binary-collision jet yield between centrality-biased AA collisions and unbiased $pp$ collisions. The different initial distributions of hard scattering vertices between the two models have little impact on the charged hadron $R_\AA$, since substantial difference only exists at large centralities where the parton energy loss is weak. On the other hand, the deviation of $R_\AA^\text{bias}$ from 1 in peripheral collisions suppresses the charged hadron $R_\AA$ in these events. The effect is sizable in the 50-70\% centrality class and is significant in 70-90\%. The suppression of charged hadrons observed in highly peripheral collisions is predominantly driven by the dilute nucleon overlap in the initial state, rather than an onset of strong jet-QGP interactions. By combining the geometric bias effect obtained from the HIJING-based model and the QGP effect simulated using the LBT model, we achieved a unified description of the charged hadron $R_\AA$ from small to large centralities in Pb+Pb collisions at $\sqrt{s_\NN}=5.02$~TeV.

This work helps improve our quantitative understanding of jet quenching in small collision systems. To further enhance the model's precision, we should develop a more solid treatment of energy depletion inside the QGP in the subsequent parton shower process and take into account additional centrality dependences of hard parton productions, such as the impact parameter dependence of the cold nuclear matter effect in the initial state~\cite{Helenius:2012wd} and the selection bias arising from centrality division~\cite{ALICE:2014xsp,Loizides:2017sqq,Park:2025mbt}. Besides, extensions of the MC-Glauber model~\cite{Ke:2025tyv} and the HIJING model~\cite{Deng:2014vda} have been formulated, which affect the initial geometry and particle spectra in small systems. A short pathlength correction to parton energy loss has been proposed in Refs.~\cite{Faraday:2023mmx,Faraday:2024qtl} and shown to reduce the hadron suppression in small systems. To facilitate future studies along this direction, our LBT model for jet quenching is publicly available at~\cite{githubLBT} and the HIJING-based initial condition model is available at~\cite{githubLBThijing}.


\vspace{6pt} 





\authorcontributions{Conceptualization and methodology, SC and CS; model development, YD, CS, and SC; data analysis, CS; validation, YD and SC; writing---original draft preparation, CS; writing---review and editing, SC, CS, and YD. All authors have read and agreed to the published version of the manuscript.}

\funding{This research was funded by the National Natural Science Foundation of China (NSFC) under Grant Nos.~12575146, 12175122, 2021-867, and~12321005.}

\dataavailability{Numerical codes for generating data used in this work are publicly available in Refs.~\cite{githubLBT,githubLBThijing}.}

\acknowledgments{We are grateful for valuable discussions with Lejing Zhang, Ran Li, Wen-Jing Xing and Guang-You Qin.}

\conflictsofinterest{The authors declare no conflicts of interest.} 



%

\section*{}

%


\reftitle{References}


\bibliography{reference_inspire}



%


\end{document}